\documentclass[letterpaper, 10 pt, journal, twoside]{IEEEtran}
\IEEEoverridecommandlockouts 
\def\BibTeX{{\rm B\kern-.05em{\sc i\kern-.025em b}\kern-.08emT\kern-.1667em\lower.7ex\hbox{E}\kern-.125emX}}
\usepackage{amsmath,amsfonts}
\usepackage[ruled]{algorithm2e}
\usepackage{array}
\usepackage[caption=false,font=normalsize,labelfont=sf,textfont=sf]{subfig}
\usepackage{textcomp}
\usepackage{stfloats}
\usepackage{url}
\usepackage{verbatim}
\usepackage{graphicx}
\usepackage{cite}
\usepackage{color,xcolor}
\usepackage {fancyhdr}

\usepackage{balance}
\usepackage{booktabs} 
\usepackage{graphics}
\usepackage{multirow}
\usepackage{amsmath}
\usepackage{threeparttable}
\usepackage{tabularx}

\begin{document}

\title{
Close-Range Human Following Control on a Cane-Type Robot with Multi-Camera Fusion}

\author{Haowen Liu$^{1}$, Fengxian Wu$^{2}$, Bin Zhong$^{1,4}$, \IEEEmembership{Student Member, IEEE}, Yijun Zhao$^{1}$, Jiatong Zhang$^{3}$,\\ Wenxin Niu$^{2}$, Mingming Zhang$^{1}$, \IEEEmembership{Senior Member, IEEE} 
\thanks{
This work has been accepted by the IEEE Robotics and Automation Letters (RA-L). Copyright have transferred.}
\thanks{This paper was recommended for publication by Editor Gentiane Venture upon evaluation of the Associate Editor and Reviewers’ comments.}
\thanks{This work was supported by the National Natural Science Foundation of China (Grant No. 62273173), the Natural Science Foundation of Shenzhen (Grant No. JCYJ20210324104203010), Shenzhen Key Laboratory of Smart Healthcare Engineering (Grant No. ZDSYS20200811144003009), the National Key Research and Development Program of China (Grant No. 2022YFF1202500, Grant No. 2022YFF1202502), the Guangdong Provincial Key Laboratory of Advanced Biomaterials (Grant No. 2022B1212010003), the Research Program of Guangdong Province (Grant No. 2020ZDZX3001, Grant No. 2019ZT08Y191).}
\thanks{$^{1}$Haowen Liu, Bin Zhong, Yijun Zhao, and Mingming Zhang are with the Department of Biomedical Engineering, Southern University of Science and Technology, Shenzhen, 518055, China.} 
\thanks{$^{2}$Fengxian Wu and Wenxin Niu are with Shanghai Yangzhi Rehabilitation Hospital (Shanghai Sunshine Rehabilitation Center), School of Medicine, Tongji University, Shanghai, 201619, China.}
\thanks{$^{3}$Jiatong Zhang is with the School of Mechanical Engineering and Automation, Harbin Institute of Technology, Shenzhen, 518055, China.}
\thanks{$^{4}$Bin Zhong is also with the Department of Biomedical Engineering, the National University of Singapore, Singapore, 117583.}
\thanks{Haowen Liu and Fengxian Wu contributed equally to this work.}
\thanks{Corresponding author: Mingming Zhang (zhangmm@sustech.edu.cn).}
\thanks{Digital Object Identifier (DOI): 10.1109/LRA.2023.3307011}

}

\markboth{IEEE Robotics and Automation Letters. Preprint Version. Accepted August, 2023}{Liu \MakeLowercase{\textit{et al.}}: Close-Range Human Following Control on a Cane-Type Robot with Multi-Camera Fusion}
\maketitle

\begin{abstract}
Cane-type robots have been utilized to assist the mobility-impaired population. The essential technique for cane-type robots is to follow the user's ambulation at a close range. This study developed a new cane-type wheeled robot and proposed a novel human-following control frame with multi-camera fusion. This human following control adopts a cascade control strategy consisting of two parts: 1) a human following position controller that locates a user by detecting his/her legs' positions via multi-camera fusion and 2) a cane robot velocity controller to steer the cane robot to the target position. The proposed strategy's effectiveness has been validated in outdoor experiments with six healthy subjects. The experimental scenarios included different terrains (i.e., straight, turning, and inclined paths), road conditions (i.e., asphalt and brick roads), and walking speeds. The obtained results showed that the average tracking error in the X and Y directions was less than 4.1 cm and 4.4 cm, respectively, and the error in angle was less than 12.9° across all scenarios. Moreover, the cane robot can effectively adapt to a wide range of individual gait patterns and achieve stable human following at daily walking speeds (0.74 m/s - 1.47 m/s).

\end{abstract}
	
\section{Introduction}

\IEEEPARstart{P}{opulation} aging has increased the burden of age-related chronic diseases, such as impaired mobility\cite{gbd2022global}. Among these mobility-impaired individuals, those with lower limb dysfunction in Category 4 or above can still walk independently according to functional ambulation classification. However, for safety considerations, they yet require a certain level of assistance during walking \cite{1984Clinical}. Thus, walking aids are commonly employed to enhance their walking independence by compensating for balance and stability\cite{jeong2015type}. 

Given the lack of nursing care staff, various robots have been developed to assist the mobility-impaired population in maintaining walking independence. These robots, commonly known as exoskeletons \cite{wang2022review,10138566}, walkers \cite{sierra2019human,ferrari2020human}, and intelligent canes \cite{fallprevention,clinicalgaittraining, clinicalgaittraining1,yan2021human}, are classified as mobility-aid robotics. Walkers typically provide support through a hand-held support base and operate by a user’s residual ambulation capability. Some solutions compensate for the walker weight and inertia\cite{li2023mobile}. However, their inherent weight and large size may limit their suitability in narrow spaces and outdoor scenarios. Additionally, they are primarily designed for individuals with severe walking dysfunction. Comparatively, cane-robot systems offer greater flexibility and smaller size, making them potentially more suitable for assisting the targeted people in various settings.

Over the past two decades, many cane-type robots have been proposed to assist and supervise mobility-impaired individuals. Pei et al. employed a cane-type walking-aid robot to estimate the risk of falls in elderly individuals, using Zero Moment Point (ZMP) stabilization as a reference for fall prevention \cite{fallprevention}. Shunki et al. used a cane-type assistive mobile robot for clinical gait training in clinical settings and verified the efficacy of their proposed gait rehabilitation strategy with the robot\cite{clinicalgaittraining, clinicalgaittraining1}. Yan et al. implemented a user intention estimation model and successfully achieved human following for the safety and supervision of users' independent walking during rehabilitation training \cite{yan2021human}. 

Among the techniques in the cane-type robots, human following control is crucial to guarantee a system's usability. The performance of human following control depends on the accuracy of the user's dynamic information (i.e., relative position and walking speed) obtained from the sensor system. 
Usually, the relative position between a cane and a human user was recommended to be 0.10 m to 0.25 m lateral to the human body \cite{article,kujath2018cane}. In this case, cane-type robots are always equipped with LiDAR sensors to acquire the user's position and walking speed \cite{algabri2020deep, Zhou2022, T2021}. For example, Huang et al. have proposed a laser-based cane-type robot to detect legs' position and follow human users during gait rehabilitation, demonstrating a satisfactory human following performance at a close range\cite{yan2021human, yan2022intelligent}. LiDARs can provide precise and high-resolution depth information for object detection, as demonstrated in studies such as \cite{yan2021human,humanleg3}. Nevertheless, they can be costly and have size, weight, and power consumption restrictions for robotic design. In addition, they may require intricate processing and calibration, particularly when used in outdoor scenarios.

In this paper, we introduced a newly designed cane-type robot with multi-camera fusion. Specifically, the cane robot detected and followed the real-time middle point position between two legs via cameras to facilitate human following. This approach was inspired by the ``point-foot" model suggesting that the middle point of two legs was highly aligned with the center of gravity during human walking\cite{garcia1998simplest,bruijn2018control}. By following this point, we are able to capture the global movement pattern, laying the foundation for comfortable human following. Cameras were preferred in this study due to the friendly price and easy setup. We applied four cameras to guarantee a sufficient detection area. With the camera system, we implemented the AprilTag-based objective detection method, which has been proven to be robust and flexible \cite{olson2011apriltag}, to obtain the user legs' positions and orientations. Additionally, to improve the robustness of tag detection during human walking, we considered occlusions of legs and established a linear regression model to compensate for the data loss caused by the occlusions, which was less discussed in human following techniques \cite{humanleg2,humanleg1}. Finally, we tested the effectiveness of the proposed cane-type robot along with the close-range human following control in the laboratory environment and various outdoor settings. The effect of the walking speed on the tracking accuracy was also investigated.

\section{Method}

\subsection{Description of the Cane Robot}

\textbf{Fig. 1} shows the configuration of the cane robot. Its sensor system includes four cameras to detect the AtrilTag attachments on both legs and a three-axis force sensor to measure the interaction force between the robot and the user's hand. Each AprilTag attachment is set to contain two AprilTags at an angle of 120°, which helps reduce the detection failure rate due to the leg's rotation during walking. The sampling rate of the camera system is 120 Hz. 
    
The wheelbase consists of four actuators (DJI M3508 DC motors with C620 drivers), four Mecanum wheels, and two 24V Lithium batteries. Notably, each wheel had a suspension structure to enhance the robot's adaptability to uneven ground surfaces. The handle on the robot is not centered but rather offset by 10 cm closer to the user.

The control system comprises a microcontroller (STM32F407) and a microcomputer (Inter NUC11 Performance Kit with Ubuntu 18.04). The human following control and data storage are run at a frequency of 120 Hz on the microcomputer. Then, the microcomputer transmits control commands to the microcontroller via a serial port. The microcontroller controls the velocity of four motors (500 Hz) and transmits the motion status of the motors and the interactive force data collected by the force sensors to the microcomputer.

\begin{figure}[tp]
\vspace{0.3cm}
 \centering
 \includegraphics[width=0.8\linewidth]{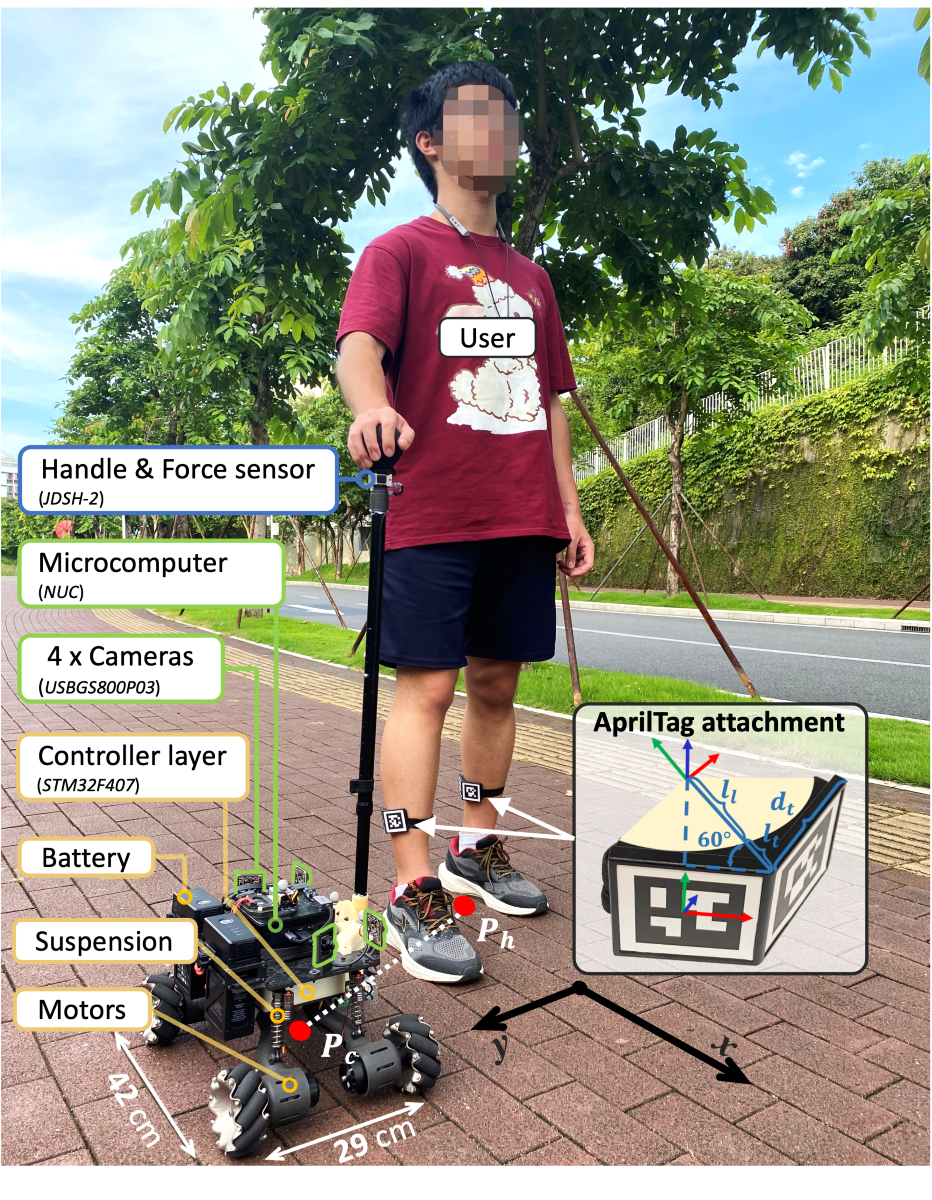}
 \vspace{-0.2cm}
 \caption{Overview of the cane robot with a human user and the configuration of the AprilTag attachment.}
  \vspace{-0.5cm}
 \end{figure}
 	
\begin{figure*}[tp]
\centering
\vspace{0.2cm}
\includegraphics[width=1\linewidth]{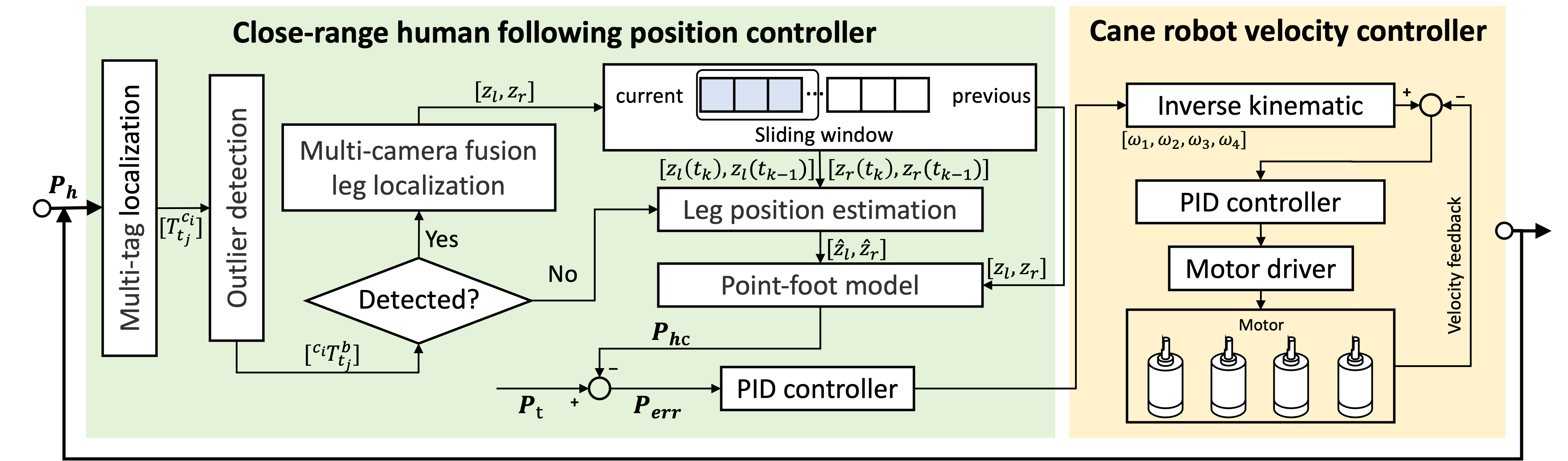}
\caption{Control frame of the proposed human-following strategy. }
\vspace{-0.5cm}
\end{figure*}
    
\subsection{Human Following Control via Multi-camera Fusion}
The control frame of the proposed human following strategy is shown in \textbf{Fig. 2}. 

In general, the strategy starts with multi-tag localization, in which the tags on both legs are tracked and located via multi-camera. Then, the outlier detection method removed the falsely identified outliers caused by lighting and complex backgrounds. After that, if one leg is detected by multiple cameras, we will use a fusion algorithm to localize the leg and record the data in a sliding window. If the leg (usually the leg farther from the robot) is not detected by any camera, we will estimate the leg's position based on the data in the sliding window. With the obtained states of both legs, the position of the human user can be calculated according to the point-foot model\cite{garcia1998simplest}. Finally, the control signal generated by the close-range human following controller is transmitted to the microcontroller to regulate the cane robot's velocity.
 
\subsubsection{Close-range human following position controller}

In multi-tag localization, it is essential to consider the failure detection of the tag. 
The pose of the $i$-th camera frame with respect to the  $j$-th tag frame can be expressed via a homogeneous transformation matrix:
\begin{equation}
    T^{ci}_{tj} = 
    \begin{cases}
     \begin{bmatrix}
    R^{ci}_{tj} & P^{ci}_{tj} \\ 
    0 & 1
     \end{bmatrix} \in \mathbb{SE}(3), & (i,j)\text{ detected} \\ 
    \mathbf{O_{4\times4}}, & \text{otherwise}
     \end{cases}
\end{equation}
where $T^{ci}_{tj}$ was determined when the $i$-th camera has detected the $j$-th tag, and $\mathbf{O_{4\times4}}$ is a null matrix when the tag was not detected. 
$R^{ci}_{tj}$ is a rotation matrix in $\mathbb{SO}(3)$ representing the relative orientation from the $j$-th tag frame to the $i$-th camera frame, and a translation vector $P^{ci}_{tj}$ in $\mathbb{R}^3$ represent the position of the $j$-th tag with respect to the $i$-th camera frame. 

Four cameras are installed as shown in \textbf{Fig. 3}. Camera 1 ($c_{0}$) and camera 4 ($c_{3}$) are installed at an angle of 30° to the robot's forwarding direction. Camera 2 ($c_{1}$) and camera 3 ($c_{2}$) are vertical to the robot's forwarding direction. On the one hand, multiple cameras increase the detection range. On the other hand, this symmetrical installation facilitates the robot's application on the user’s both sides. 

However, due to the complexity of outdoor environments, the absolute 6D pose data obtained from tag recognition may result in outliers. In order to exclude such outliers, a ``Detection Region" (see \textbf{Fig. 3}) for the tags' positions has been defined based on the range between the user and the cane robot. Tag positions outside this region are treated as outliers and excluded. The pose of the $j$-th tag relative to the cane body frame, as captured by the $i$-th camera, can be represented by
\begin{equation}
^{ci}T^{b}_{tj}=  
\begin{cases}
T^{b}_{ci}T^{ci}_{tj},
& \mbox{if } ^{ci}P^{b}_{tj}  \in R_1
\\ \mathbf{O_{4\times4}}, 
& \text{otherwise}
 \end{cases}
\end{equation}
where $T^{b}_{ci}(t) \in \mathbb{SE}(3)$ is a homogeneous transformation matrix denoting the relative pose from the $i$-th camera frame to the cane body frame. 
$^{ci}P^{b}_{tj}$ represents the position of the $j$-th tag relative to the cane robot, as captured by the $i$-th camera. After removing the outliers, we obtained the pose of multiple tags captured by different cameras. 

The pose relationship between the leg and tags (see \textbf{Fig. 1}) can be expressed as a homogeneous transformation matrix as follows:
\begin{equation}
\begin{aligned}
T_{l}^{t0}=T_{r}^{t2}=
\small{
\begin{bmatrix}
0.866& -0.5& 0& 0.5(l_l+l_t) - d_t
\\ 0& 0& -1& 0
\\ 0.5& 0.866& 0&\frac{\sqrt{3}}{2}(l_l+l_t)
\\ 0& 0& 0& 1
\end{bmatrix}
}
\end{aligned}
\end{equation}
\begin{equation}
\begin{aligned}
T_{l}^{t1}=T_{r}^{t3}=
\small{
\begin{bmatrix}
0.866& 0.5& 0& -0.5(l_l+l_t) + d_t
\\ 0& 0& -1& 0
\\ -0.5& 0.866& 0&\frac{\sqrt{3}}{2}(l_l+l_t)
\\ 0& 0& 0& 1
\end{bmatrix}
}
\end{aligned}
\end{equation}
where $l_l$ is the radius of the calf, and $l_t$ (22.5 mm), $d_t$ (50 mm) are size parameters for the AprilTag attachment. 

The states of the left and right legs are defined as follows:
\begin{equation}
    \begin{aligned}
&z_l=[x_l,y_l,\theta_l]^T
\\ &z_r=[x_r,y_r,\theta_r]^T
    \end{aligned}
\end{equation}
where $x_l$, $y_l$, and $\theta_l$ are the measured user’s leg position and orientation in the cane robot's coordinate system. 

We proposed an algorithm for improving the accuracy of leg localization by using mean filtering to process multiple data of the same tag. 

If any tag corresponding to a leg is recognized, we can locate the state of the leg. For example, the state of the left leg is calculated as below (the state of the right leg is similar):
\begin{equation}
x_l=\frac{\sum_{i=0}^3\sum_{j=0}^1 L_1({^{ci}R^{b}_{tj}P^{tj}_{l}}+{^{ci}P^{b}_{tj})}}
{\sum_{i=0}^3\sum_{j=0}^1G(^{ci}T^{b}_{tj}T^{tj}_{l})}
\end{equation}
\begin{equation}
			y_l=\frac{\sum_{i=0}^3\sum_{j=0}^1 L_2({^{ci}R^{b}_{tj}P^{tj}_{l}}+{^{ci}P^{b}_{tj})}}
{\sum_{i=0}^3\sum_{j=0}^1G(^{ci}T^{b}_{tj}T^{tj}_{l})}
\end{equation}
\begin{equation}
            \theta_l=\frac{\sum_{i=0}^{3}\sum_{j=0}^1 \mathrm{J}_\Theta({^{ci}R^{b}_{tj}R^{tj}_{l}})}
{\sum_{i=0}^3\sum_{j=0}^1G(^{ci}T^{b}_{tj}T^{tj}_{l})}\;
\end{equation}
where 
\begin{equation}
G(T)=\begin{cases}
1, & \mbox{if } T \ne \mathbf{O_{4\times4}}
\\ 0, & \text{otherwise}
\end{cases}
\end{equation}$L_1=[1,0,0]$, $L_2=[0,1,0]$, the numbering of cameras and tags start from 0 and $\mathrm{J}_\Theta$ represents the conversion operation from rotation matrix to Euler angle\cite{Euler}. 

Then, the state of the legs is stored in a sliding window for leg pose estimation. We refer to the timings of the two previous detections of the leg as $t_{k}$ and $t_{k-1}$, respectively. In leg pose estimation, if all tags located on one leg are missed at time $t_c$, The time intervals are calculated as:
\begin{equation}
    \begin{aligned}
      &\Delta T_1=t_{c}-t_{k}
      \\ &\Delta T_2=t_{k}-t_{k-1}
    \end{aligned}
\end{equation}
The undetected leg’s pose $z(t_c)$ can be estimated by previous detections of the leg based on the following rule-based method:
\begin{equation}
\hat{z}_l(t_{c})=  
\begin{cases}
z_r(t_{k-1})+[0,d,0]^T,
& \Delta T_1>\delta_1
\\ z_l(t_{k-1})+k(\Delta T_1),
& \Delta T_1<\delta_1 \text{ and } \Delta T_2<\delta_2
\\ z_l(t_{k-1}),
& \text{Otherwise}
 \end{cases}
\end{equation}
where $k=(z_l(t_{k})-z_l(t_{k-1}))/(t_{k}-t_{k-1})$. $d$ represents the distance between both legs in the vertical direction (y-axis) while standing, and $d$ is set to be 20 cm based on empirical observations. $\delta_1$ and $\delta_2$ are the threshold values used to determine whether an individual has come to a stop and whether linear estimation should be employed, respectively.  $[\delta_1$, $\delta_2]$ are set to be [100, 60] based on the user's walking speed and spatiotemporal parameters of our experiment.

After obtaining the states of both legs, the position of the human user $P_{hc}$ is calculated by the ``point-foot model" as:
\begin{equation}
P_{hc}=[x_{hc}, y_{hc},\theta_{hc}]^T=0.5[x_r+x_l, y_r+y_l , \theta_r+\theta_l]^T
\end{equation}

During  human walking, the change of the $P_{hc}$ is characterized by a low-frequency signal\cite{winter1995human}. Thus, we added a Butterworth low-pass filter with a cutoff frequency of 100 Hz to enhance the stability of the system.

\subsubsection{Cane robot velocity controller}
 After obtaining the user's position, we set the desired position of the cane robot as control inputs, and then we use the PID controller to control the position of the cane robot. We calculated the inverse kinematic relationship between the cane robot's velocity in each direction and the motor's rotational speed in accordance with the method presented in\cite{Mecanum }. The relationship is as follows:
\begin{equation}
    \begin{bmatrix}
    w_1\\  w_2 \\  w_3 \\w_4
    \end{bmatrix}
    =\frac{1}{r}
    \begin{bmatrix}
      1&-1  &(a+b)
    \\1&1 &-(a+b)
    \\1&-1  &-(a+b)
    \\1&1  &(a+b)
    \end{bmatrix}
    \begin{bmatrix}
    v_x\\  v_y \\  w_z 
    \end{bmatrix}
\end{equation}where $r$ is the radius of  the Mecanum wheel, and $[w_1,w_2,w_3,w_4]^T$ is the angular velocity of the motor. $v_x,v_y$ is the velocity of the cane robot in the \textit{x-y} direction, and $w_z$ is the angular velocity of the robot. 
\begin{figure}[!tp]
\vspace{0.3cm}
 \centering
 \includegraphics[width=0.9\linewidth]{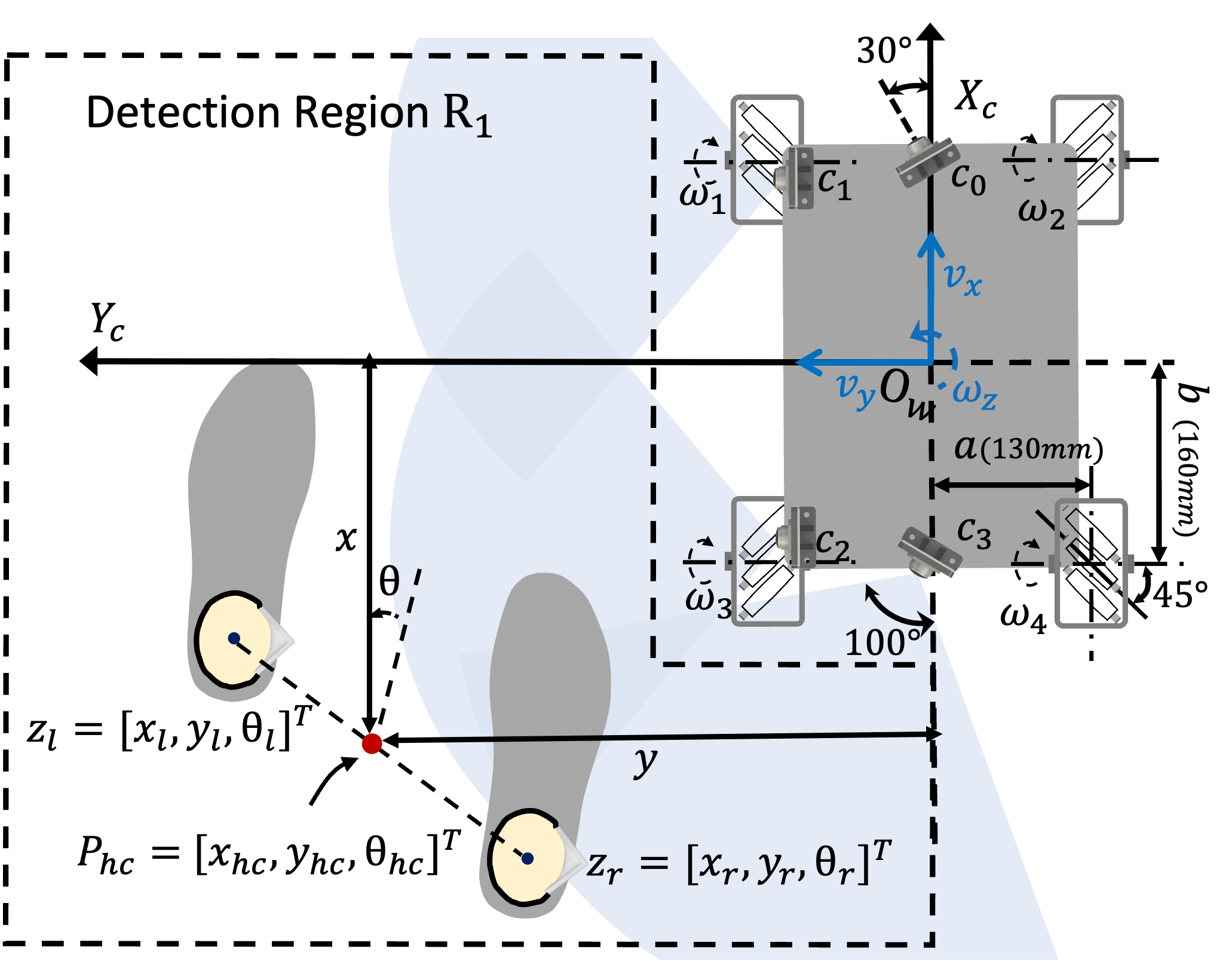}
 \vspace{-0.1cm}
 \caption{Illustration of the cameras' installation and the definition of the cane robot's detection region.}
 \vspace{-0.5cm}
\end{figure}
\subsection{Experimental Validation of the Cane Robot}

\begin{table*}[ht]
\vspace{0.5cm}
    \caption{IDENTIFICATION OF PARTICIPANTS’ CHARACTERISTICS}
    \vspace{-0.2cm}
    \tiny 
    \resizebox{\linewidth}{!}{
        \centering
       \tiny 
        \begin{tabular}{cccccccccccccc}
        \toprule 
        \multirow{2}{*}{\textbf{Participant}} & \multirow{2}{*}{\textbf{Sex}} & {\textbf{Age}} & {\textbf{Height}}  & {\textbf{Weight}} & \multicolumn{2}{c}{\textbf{Speed in Exp 1 (m/s)}}  & \multicolumn{7}{c}{\textbf{Speed in Exp 2 (m/s)}}  \\ \cmidrule[0.15mm](lr){6-7} \cmidrule[0.15mm](lr){8-14}
        & & (years) & (m) & (kg) &path 1 &path 2 & 1 & 2 & 3 & 4 & 5 & 6 & 7 \\
        \midrule 
              1 & M & 24 & 1.70 & 65 & 1.02 & 1.02 & 0.75 & 0.93 & 0.94 & 1.06 & 1.10 & 1.30 & 1.40  \\ 
              2 & M & 23 & 1.73 & 69 & 1.02 & 1.02 & - & - & - & - & - & -& - \\ 
              3 & F & 24 & 1.60 & 50 & 1.08 & 1.06 & 0.84 &	0.97 &	1.00 &	1.07 &	1.10 &	1.37 &	1.47 \\ 
              4 & M & 24 & 1.72 & 56 & 1.03 & 0.96 & -& & -&- &- & - &-\\ 
              5 & M & 26 & 1.70 & 55 & 1.00 & 1.00 &- &- & -& -& -&- &-\\ 
              6 & F & 26 & 1.64 & 60 & 0.82 & 0.82 & 0.74 & 0.82 & 0.97 &	1.07 & 1.19 &  1.30 & 1.39\\ 
        \bottomrule 
        \end{tabular}
     }
     \begin{tablenotes}
    \footnotesize
    \item[]\textbf{*Note:} "Speed in Exp 1" is "Average walking speed in experiment 1". "Speed in Exp 2" is "Average walking speed in experiment 2".
      \end{tablenotes}
      \vspace{-0.5cm}
\end{table*}

This study focused on validating the feasibility and efficacy of the cane robot with close-range human following control. This study involved human participants and was approved by the Southern University of Science and Technology, Human Participants Ethics Committee (No.20220030).
    
\subsubsection{Evaluation of system's performance}
We carried out three tests to investigate the system's performance as follows:
 
\textbf{Test 1}: We evaluated the velocity tracking performance of the robot by following sinusoidal velocity profiles on a flat road surface. For the profiles, we selected three magnitudes (corresponding to daily walking speeds: 0.5 m/s, 1.0 m/s, and 1.5 m/s) and frequencies at 0.2 Hz and 0.5 Hz. 

\textbf{Test 2}: We evaluated the position tracking performance of the cane robot by tracking the fixed AprilTag at a series of velocities in both the X and Y directions on a treadmill (see \textbf{Fig. 4(B)}). Human walking exhibits a wider range of speeds in the forward (X) direction compared to the lateral (Y) direction. Therefore, the velocity range in the X direction was from 0.5 m/s to 1.5 m/s (13 trials), while the velocity range in the Y direction was from 0.25 m/s to 0.75 m/s (7 trials). The actual positions between the robot and AprilTag were obtained by an optical motion capture system simultaneously.

\textbf{Test 3}: We performed validation tests to assess the accuracy of camera-based human motion detection using the optical motion capture system for comparison. The setup of this test is shown in \textbf{Fig. 4(C)}, and the subject walked on a treadmill at speeds from 0.5 m/s to 1.5 m/s (5 trials), while the cane robot was positioned in front of the treadmill to detect the position of the human. We validated camera accuracy by comparing the detected position between the legs' midpoint and the fixed AprilTag with the distance obtained from the optical motion capture system.

\subsubsection{Human-following performance in outdoor scenarios}
The human testing included six healthy subjects without joint injuries (four males and two females, age 24.3 ± 0.94 years, height 1.68 ± 0.05 m, weight 59.17 ± 6.36 kg). The individual specifications are listed in \textbf{Table I}, and all participants provided informed consent before experiments. During the experiments, the investigators can stop the cane robot immediately via a remote controller if unexpected situations occur.

Considering future application scenario of our cane robot (i.e., community walking assistance), we performed the human following validation experiments with daily terrains (i.e., straight, turning, and inclined paths), and two common types of ground surfaces were included: asphalt roads and brick roads.
A 176 m outdoor ground path was chosen as the site of the experiment (see \textbf{Fig. 5(A)}), which consists of a mild incline section at the beginning (approx. 46 m), a mild turning section at the middle (approx. 57 m), a relatively straight section at the end (approx. 73 m).  \textbf{Fig. 5(B)} illustrates the experimental protocol as follows:

\textbf{Experiment 1}: the first experiment aimed to evaluate the robot's following performance on common terrains. Six healthy adults participated in this experiment. Before the walking trial, we measured the participants' lower limb sizes and adjusted the tags at the same height as the camera to achieve better leg detection. Then, the participants walked 2-min with the robot following for familiarization. Subsequently, the participants were instructed to complete two walking tests in two paths from A to D as shown in \textbf{Fig. 5(A)}. A 5-min break was given between each trial to prevent potential fatigue effects.

\textbf{Experiment 2}: the second experiment focused on the feasible walking speed range of the robot. Three subjects (i.e., P1, P3, and P6) participated in this experiment. The participants first adjusted the tags as described in Experiment 1. Subsequently, each participant completed seven trials by
walking from A to D in self-selected speeds. They were instructed to cover a range of daily walking speeds. A 5-min break was given between each trial.
    
During all the tests, the robot was programmed to keep a fixed position to the user. The fixed relative posture of the human-following rule is determined as $[x,y,z]=[350,450,0]$ (see \textbf{Fig. 3}) based on the clinical recommendations for using a unilateral cane \cite{kujath2018cane} and the principles of proxemics theory in human following robot\cite{rios2015proxemics}. 

\subsection{Statistics}
The velocity tracking performance of the robot was evaluated by the error between the desired velocity profiles and actual velocity profiles across five sinusoidal cycles. In experiment 1, the average tracking error was calculated as the absolute deviation between the robot's target position and its actual position and orientation, which was indicated by the values of $\left|emx\right|$, $\left|emy\right|$, and $\left|em\theta\right|$. Additionally, a Two-Way Repeated Measures ANOVA was conducted to determine the impact of ground surfaces (asphalt and brick roads) and terrain (incline, turning, and straight paths) on average tracking error, with Shapiro-Wilk test and Mauchly's spherical hypothesis test performed (a significance level of alpha = 0.05). Lastly, in experiment 2, the average tracking error was quantified by calculating the Euclidean distance difference between the target and actual positions. A simple linear regression model was used to analyze the correlation between walking speed on tracking accuracy.

 \section{Results}
 
\subsection{Tracking Performance of the Cane Robot}
\textbf{Fig. 4(A)} shows the average velocity tracking errors of the robot in the X and Y directions and rotation for specific sinusoidal velocity profiles. The results demonstrated that increasing the velocity and frequency led to higher tracking errors in the X and Y directions and rotation. The highest error was recorded when the maximum velocity was 1.5 m/s, and the frequency was 0.5 Hz, resulting in mean errors of 0.107 m/s for X, 0.105 m/s for Y, and 0.246 rad/s for rotation, respectively. The results suggested that the robot effectively achieved accurate velocity tracking during daily walking, as the typical walking speed during walking is within 1.5 m/s according to previous studies \cite{waters1999energy}.

\textbf{Fig. 4(B)} presents the average position tracking errors of the robot in X and Y directions and rotation for various x and Y direction treadmill velocities. The results demonstrated that increasing the velocity led to higher tracking errors in X and Y directions and rotation. Error in Y direction is smaller than error in the X direction, and there is no significant increase in error in the Y direction as the speed increases, whether the treadmill velocity is in X or Y directions. Overall, the average error in the X direction is below 30 mm, the average Y error is below 14 mm, and the average error in angle is below 4°.

\textbf{Fig. 4(C)} demonstrates that our system consistently detects human positions in the range of typical human walking speeds. Compared to the optical motion capture system, the average error in the X direction is below 5.4 mm, the average error in the Y direction is below 2.4 mm, and the average angle error is below 2.1°.

\begin{figure}[!htp]
 \vspace{0.4cm}
 \centering
 \includegraphics[width=1\linewidth]{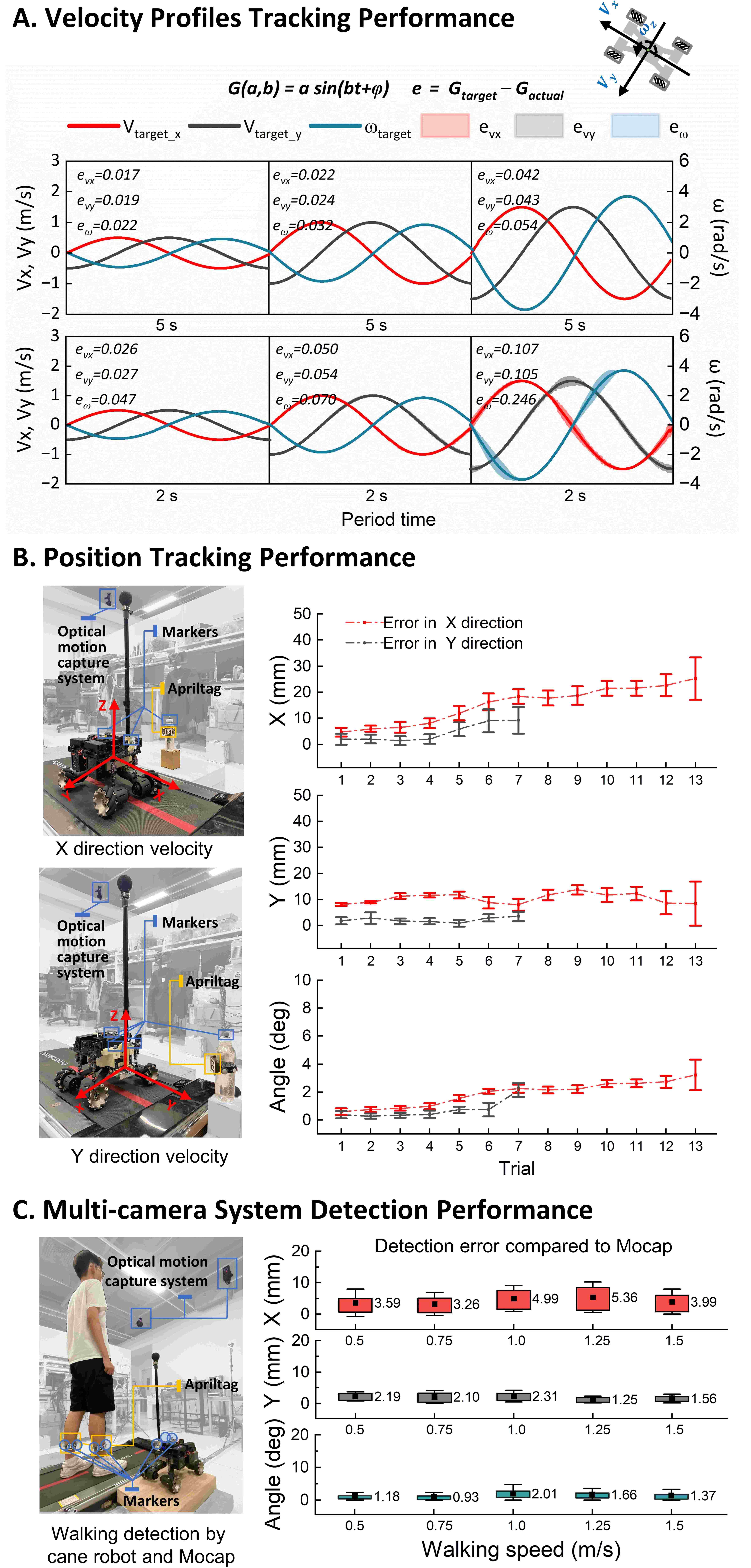}
 \vspace{-0.5cm}
 \caption{The setup and corresponding results of validation tests. “Mocap” is “Optical motion capture system ”} 
 \vspace{-0.7cm}
\end{figure}

\subsection{Human-following Performance in Outdoor Scenarios}
\textbf{Fig. 5(C)} illustrates the cane robot following rules and the 10-second real-time human following error of six participants. During the real-time following, the cane robot exhibits fluctuations around the target position. Furthermore, through analysis of individuals, we observed that the distribution of errors was associated with the gait period. Specifically, the error increased during the push-off and middle-stance due to the rapid change in calf direction during the push-off and the being obstructed during the middle-stance.  

The average tracking error of $\left|emx\right|$, $\left|emy\right|$, and $\left|em\theta\right|$ of six participants during walking on two grounds and three terrains are depicted in \textbf{Fig. 5(D)}. The average values for $\left|emx\right|$ and $\left|emy\right|$ are less than 4.1 cm and 4.4 cm, respectively, and the average value for $\left|em\theta\right|$ is less than 12.9° across all scenarios. Two-Way Repeated Measures ANOVA results show that no significant interaction was found through post-hoc analysis of the ground and terrain factors ($\left|emx\right|$: $F$=0.52, $p$=0.608; $\left|emy\right|$: $F$=0.75, $p$=0.497; $\left|em\theta\right|$: $F$=1.70, $p$=0.231). Thus, the main effects of the two within-subject factors (ground and terrain) were analyzed. The results show that no statistical significance was found with regards to the effect of terrain and ground factors on the following error for $\left|emx\right|$, $\left|emy\right|$, and $\left|em\theta\right|$.

To evaluate the range of achievable walking speeds for the robot human-following, we conducted an experiment in which three participants walked at seven different speeds ranging from 0.74 m/s to 1.47 m/s, covering daily walking speeds (see \textbf{Fig. 5(E)}). The tracking error distributions of the three participants are similar, with Participant 3 exhibiting an average tracking error of 58.1 mm, which is slightly larger than that of Participants 1 and 2 (45.0 mm and 46.3 mm, respectively). Additionally, the correlation analysis reveals that there is no statistically significant correlation between the walking speed and the tracking error on the incline and straight sections, as indicated by $R^{2}$=0.23 and $R^{2}$=0.16, respectively, both $p$$>$ 0.05. However, the following error of the turning section was found to decrease as the walking speed increased, as demonstrated by $R^{2}$=0.34, $p$$<$0.05.
\section{Discussion}
This study presented a novel control scheme for a cane-type robot that achieves stable and accurate close-range human following in outdoor scenarios. Notably, all participants completed the experiment without any interference from the robot, indicating that the robot can be effectively adapted to a wide range of individual gait patterns.  
\subsection{Dynamic following Performance}
We achieved high motion capture accuracy by establishing the detection range and implementing a multi-camera mean filtering method. Moreover, the velocity and position tracking results show that the proposed cascade PID controller achieved stable and accurate human following. However, the dynamic following accuracy of the cane robot is influenced by individual variations in walking behavior. More specifically, the highest tracking errors occurred during the push-off and middle-stance phases of the gait cycle, caused by rapid changes in lower leg direction and tag obstruction, respectively. The dynamic following performance can be improved if the walking behavior induced following error can be compensated properly in future work.
\begin{figure*}[tp]
 \centering
 \includegraphics[width=1\linewidth]{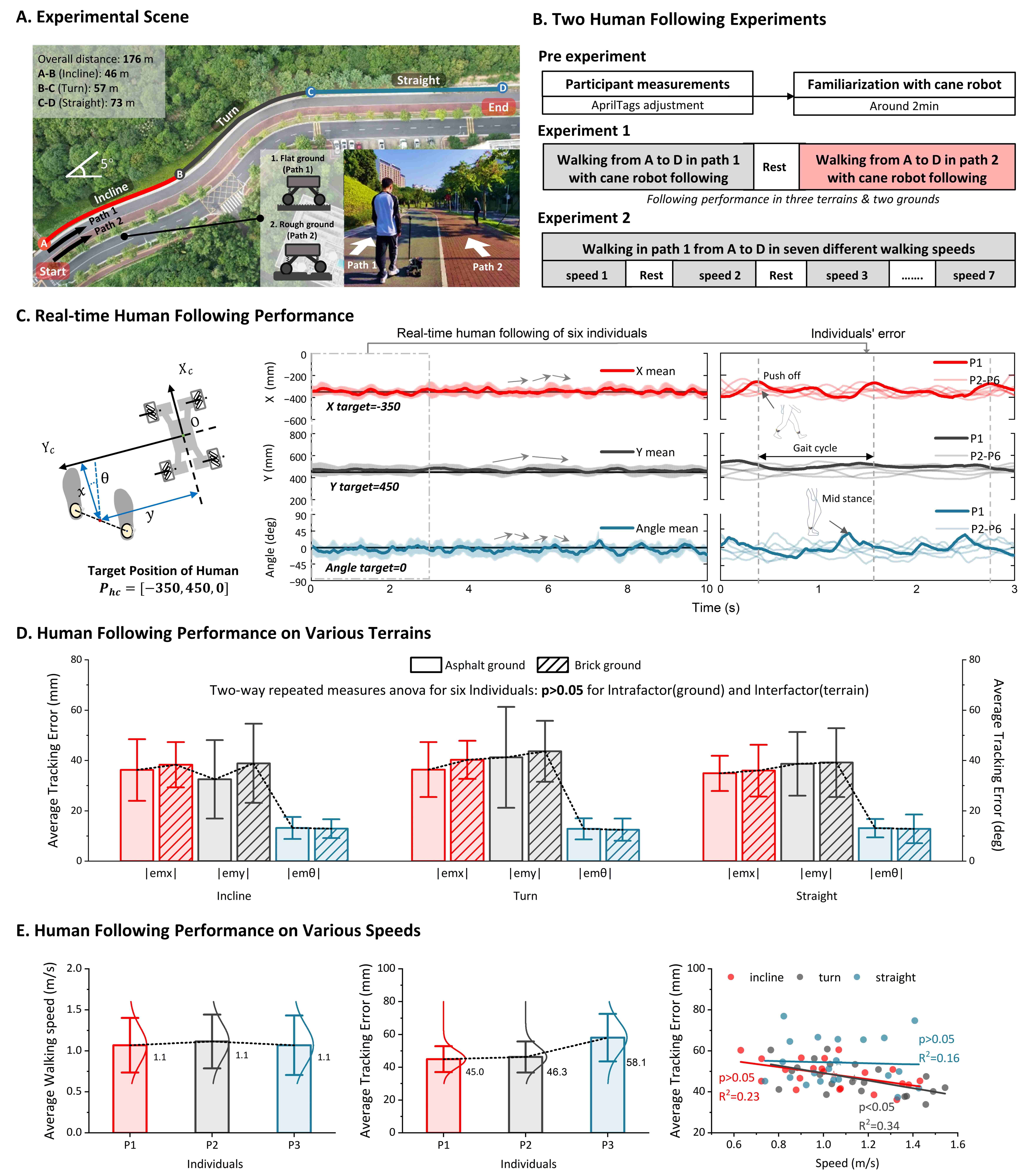}
 \caption{Experimental setup and results of human tests. A. The experimental scene includes three terrains, i.e., the incline, turning, and straight, with two kinds of common ground surfaces(asphalt and brick roads). B. Illustration of two human-following experiment protocols. C. Example of 10s X, Y, and angle tracking error curve of six participants. D. Mean and standard error result of tracking error of six participants under three terrains and two grounds. A two-way ANOVA was conducted to analyze the effect of terrain and ground factor. E. Effects of walking speed on human-following performance. Mean and standard error of three participants' average walking speed, average tracking error, and correlation analysis of speed and error.}   
\end{figure*}  
\subsection{Feasible Range of Walking Scenarios}
It is widely acknowledged that complex outdoor scenarios can significantly affect the human following performance of the cane robot. We investigated the effect of different outdoor scenarios on the tracking performance of the robot and found that the average tracking error was not statistically significant under different terrains and ground surfaces. We attribute this to two main factors: multi-camera fusion and the design of the robot's structure. Specifically, the multi-camera fusion method reduced detection errors and increased detection reliability, and the springs on the robot's legs provided shock absorption and reduced the influence of uneven ground on the robot's performance. However, further investigation is required to determine the specific contribution of each factor to the robot's performance in complex outdoor scenarios.

Additionally, our study shows that the robot achieved stable tracking without a significant correlation between tracking error and speed. This contradicts our initial hypothesis, as velocity tracking error typically increases with speed. We speculate that the reduction of occlusion at higher speeds may be a contributing factor, but further research is necessary to confirm this.

Lastly, it is difficult to directly compare the performance of different robotic systems due to several factors such as different functional scopes, hardware used, and  Validation settings. However, tracking error is the direct embodiment of human following performance and our human-following performance is comparable to the published paper\cite{humanfollowing}.

\subsection{Limitation}
This study is also subject to three limitations: 1) human position recognition relies entirely on the AprilTags attachment on the calf. If the AprilTags are covered by clothing such as a skirt or wide pants, it can lead to detection failure; 2) the effect of individual walking behaviors on the human following performance was not considered; 3) while the cane robot has shown satisfactory performance on the asphalt and brick roads, its adaptability to tough road conditions like gravel has not been verified.
\section{Conclusion}
In conclusion, this study presented a novel human following control with multi-camera fusion on a cane-type robot during outdoor walking. We utilized an AprilTag-based detection method to track the position of the legs and applied four cameras to ensure a sufficient detection scope. The sensor system was cost-friendly. Notably, by tracking the middle point between two legs based on the point-foot model, our system achieved an effective human following performance under common walking scenarios. Moreover, we considered the occlusions of legs during human walking and established a linear regression model to compensate for the data loss. The results demonstrated the effectiveness of our proposed close-range human following strategy in outdoor environment. Future studies could explore the potential of integrating machine learning algorithms to optimize the human following performance with adaptive human-robot relative distance. We also intend to integrate the cane robot with other assistive devices, e.g., lower limb exoskeletons, to achieve advanced human-robot interaction in clinical settings. 

\bibliographystyle{IEEEtran}   
\bibliography{ref}  

\end{document}